\documentclass[letterpaper,twocolumn,prl,aps,superscriptaddress,amsmath,amssymb,floatfix]{revtex4-1}
\usepackage{mathptmx}
\usepackage[latin9]{inputenc}
\usepackage{color}
\usepackage{amsmath}
\usepackage{amssymb}
\usepackage{graphicx}
\usepackage{esint}
\usepackage[unicode=true,
 bookmarks=true,bookmarksnumbered=false,bookmarksopen=false,
 breaklinks=false,pdfborder={0 0 1},backref=false,colorlinks=true]
 {hyperref}
\hypersetup{
 linkcolor=magenta,urlcolor=blue,citecolor=blue,pdfstartview={FitH},hyperfootnotes=false}

\makeatletter

\pdfpageheight\paperheight
\pdfpagewidth\paperwidth

\usepackage{textcomp}
\usepackage{epstopdf}

\pdfpageheight\paperheight
\pdfpagewidth\paperwidth

\@ifundefined{textcolor}{}{%
 \definecolor{BLACK}{gray}{0}
 \definecolor{WHITE}{gray}{1}
 \definecolor{RED}{rgb}{1,0,0}
 \definecolor{GREEN}{rgb}{0,1,0}
 \definecolor{BLUE}{rgb}{0,0,1}
 \definecolor{CYAN}{cmyk}{1,0,0,0}
 \definecolor{MAGENTA}{cmyk}{0,1,0,0}
 \definecolor{YELLOW}{cmyk}{0,0,1,0}
}

\usepackage{xcolor}\usepackage{soul}
\definecolor{blue}{rgb}{0,0,1}
\definecolor{red}{rgb}{1,0,0}
\definecolor{green}{rgb}{0,1,0}

\makeatother

\begin{document}
\title{Efficient frequency conversion in a degenerate $\chi^{(2)}$ microresonator}
\author{Jia-Qi~Wang}
\thanks{These two authors contributed equally to this work.}
\affiliation{Key Laboratory of Quantum Information, Chinese Academy of Sciences,
University of Science and Technology of China, Hefei 230026, P. R. China.}
\affiliation{CAS Center For Excellence in Quantum Information and Quantum Physics,
University of Science and Technology of China, Hefei, Anhui 230026,
P. R. China.}
\author{Yuan-Hao~Yang}
\thanks{These two authors contributed equally to this work.}
\affiliation{Key Laboratory of Quantum Information, Chinese Academy of Sciences,
University of Science and Technology of China, Hefei 230026, P. R. China.}
\affiliation{CAS Center For Excellence in Quantum Information and Quantum Physics,
University of Science and Technology of China, Hefei, Anhui 230026,
P. R. China.}
\author{Ming~Li}
\affiliation{Key Laboratory of Quantum Information, Chinese Academy of Sciences,
University of Science and Technology of China, Hefei 230026, P. R. China.}
\affiliation{CAS Center For Excellence in Quantum Information and Quantum Physics,
University of Science and Technology of China, Hefei, Anhui 230026,
P. R. China.}
\author{Xin-Xin~Hu}
\affiliation{Key Laboratory of Quantum Information, Chinese Academy of Sciences,
University of Science and Technology of China, Hefei 230026, P. R. China.}
\affiliation{CAS Center For Excellence in Quantum Information and Quantum Physics,
University of Science and Technology of China, Hefei, Anhui 230026,
P. R. China.}
\author{Joshua B.~Surya}
\affiliation{Department of Electrical Engineering, Yale University, New Haven,
CT 06511, USA}
\author{Xin-Biao~Xu}
\affiliation{Key Laboratory of Quantum Information, Chinese Academy of Sciences,
University of Science and Technology of China, Hefei 230026, P. R. China.}
\affiliation{CAS Center For Excellence in Quantum Information and Quantum Physics,
University of Science and Technology of China, Hefei, Anhui 230026,
P. R. China.}
\author{Chun-Hua~Dong}
\affiliation{Key Laboratory of Quantum Information, Chinese Academy of Sciences,
University of Science and Technology of China, Hefei 230026, P. R. China.}
\affiliation{CAS Center For Excellence in Quantum Information and Quantum Physics,
University of Science and Technology of China, Hefei, Anhui 230026,
P. R. China.}
\author{Guang-Can~Guo}
\affiliation{Key Laboratory of Quantum Information, Chinese Academy of Sciences,
University of Science and Technology of China, Hefei 230026, P. R. China.}
\affiliation{CAS Center For Excellence in Quantum Information and Quantum Physics,
University of Science and Technology of China, Hefei, Anhui 230026,
P. R. China.}
\author{Hong X.~Tang}
\affiliation{Department of Electrical Engineering, Yale University, New Haven,
CT 06511, USA}
\author{Chang-Ling~Zou}
\email{clzou321@ustc.edu.cn}

\affiliation{Key Laboratory of Quantum Information, Chinese Academy of Sciences,
University of Science and Technology of China, Hefei 230026, P. R. China.}
\affiliation{CAS Center For Excellence in Quantum Information and Quantum Physics,
University of Science and Technology of China, Hefei, Anhui 230026,
P. R. China.}
\date{\today}
\begin{abstract}
Microresonators on a photonic chip could enhance nonlinear optics
effects, thus are promising for realizing scalable high-efficiency
frequency conversion devices. However, fulfilling phase matching conditions
among multiple wavelengths remains a significant challenge. Here,
we present a feasible scheme for degenerate sum-frequency conversion
that only requires the two-mode phase matching condition. When the
drive and the signal are both near resonance to the same telecom mode,
an efficient on-chip photon-number conversion efficiency upto $42\%$
was achieved, showing a broad tuning bandwidth over $250\,\mathrm{GHz}$.
Furthermore, cascaded Pockels and Kerr nonlinear optical effects are
observed, enabling the parametric amplification of the optical signal
to distinct wavelength in a single device. The scheme demonstrated
in this work provides an alternative approach to realizing high-efficiency
frequency conversion and is promising for future studies on communications,
atom clocks, sensing and imaging.
\end{abstract}
\maketitle
\emph{Introduction.-} Coherent frequency conversion processes between
distinct frequency bands are necessary for a number of classical and
quantum optical applications~\citep{Boyd2003,Fejer1994,Caspani2011,ArunKumar2013},
ranging from communication, detection, sensing, to imaging~\citep{Curtz2010,Gottesman2012,Li2017,Guo2020}.
For these applications, we wish to convert the frequencies of the
encoded carriers while leaving the information essentially unperturbed.
Nonlinear optical processes achieve this goal by compensating for
the energy differences between the carriers while also conserving
the encoded information~\citep{Kumar1990}. In the past decades,
such coherent frequency conversion processes have been experimentally
verified~\citep{Huang1992,McGuinness2010,Clemmen2016,Rutz2017} and
have been applied in a wide range of devices, such as up-converted
single photon detectors~\citep{Vandevender2004,Samblowski2014},
long-distance entanglement distribution~\citep{Maring2017,Yu2020},
and microwave-to-optical transducers~\citep{Andrews2014,Williamson2014,Fan2018,Han2020,Rueda2016}.

Recently, with the improvement of nanofabrication technology, integrated
nonlinear photonic chips have made significant technological advances
while increasingly drawing more attention from the scientific community~\citep{PRL_2013_zeno,Strekalov2016,Breunig2016,Jiang2018,Xie2019}.
These integrated photonic chips have shown better performances than
macroscopic waveguides and resonators because of the strong confinement
of optical fields and the long lifetime of the optical resonances,
which allow high conversion efficiency, miniaturized footprint, excellent
scalability, and potential to integrate with other on-chip components.
For example, high-efficiency frequency conversion and strong coupling
between optical modes in distinct wavelengths have recently been realized
by microring resonators~\citep{Guo2016a,Li2016,PRL_SiN_strong_couple,LuXiyuan2019}.
Additionally, materials with strong nonlinearity ($\chi^{(2)}$),
such as LiNbO$_{3}$, AlN, GaAs and AlGaAs~\citep{Lu2019,ChenLN2019,Lu2020,Kuo2014,Chang2018,Stanton2020,Chang2019},
have transitioned on-chip for achieving even larger photon-photon
interaction and other promising qubit-free optical devices~\citep{Li2020,Heuck2019,Krastanov2020}.
Along with the advantages of stability and scalability, the photonic
chip also holds great potential for future quantum information processing
and quantum network applications~\citep{Wang2019,Elshaari2020}.
The main challenge for resonant-enhanced coherent frequency conversion
is the constraints of simultaneously satisfying energy and momentum
conservation for multiple modes at distinct wavelengths, especially
for matching the target transitions of atoms or other emitters.\textcolor{red}{{}
}
\begin{figure}
\begin{centering}
\includegraphics[width=1.02\columnwidth]{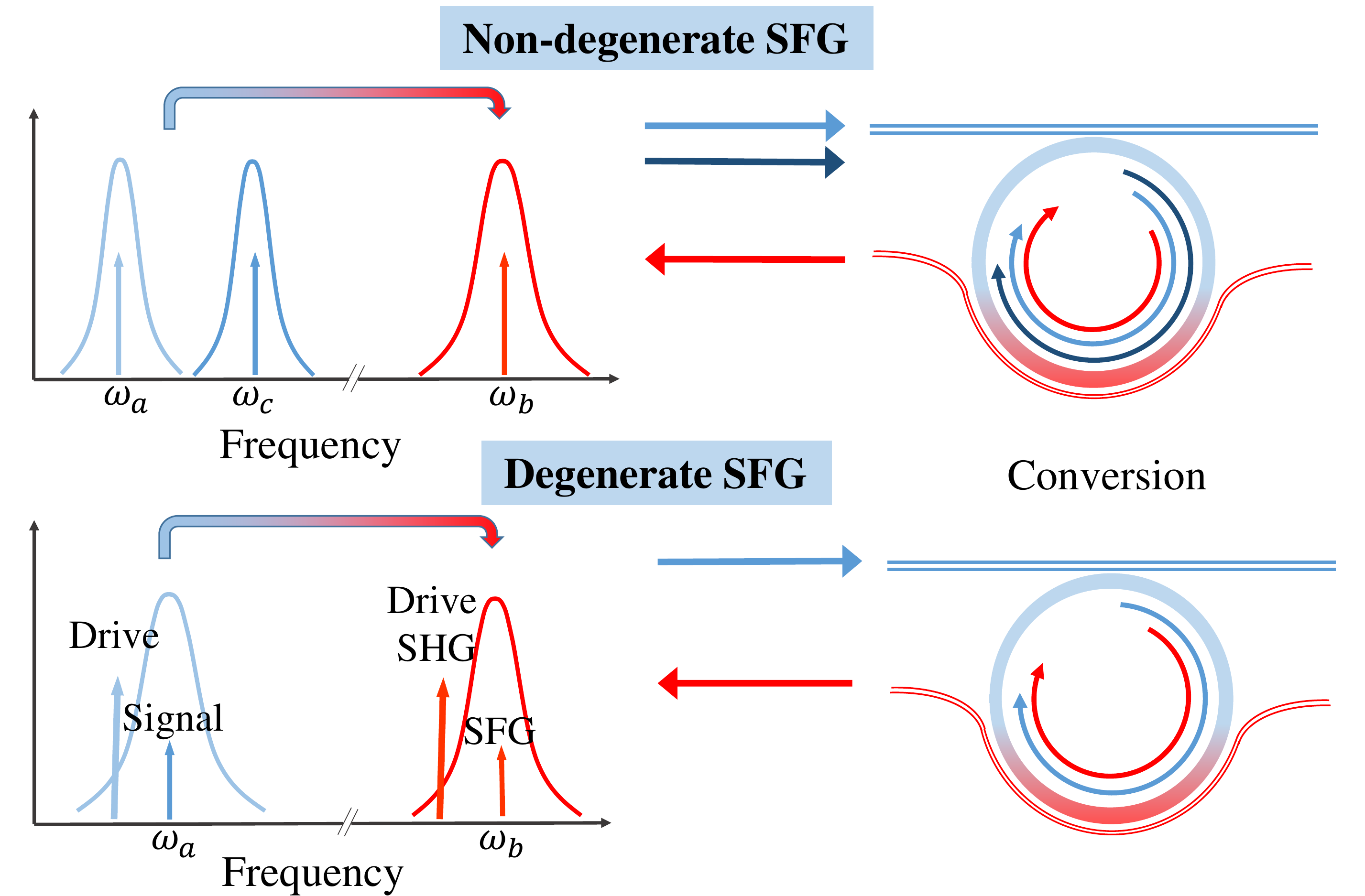}
\par\end{centering}
\caption{Schematic illustration of the frequency conversion via non-degenerate
sum-frequency generation and degenerate sum-frequency generation models.
Comparing the two schemes, the degenerate case only requires the phase
matching condition between two modes and produces a second-harmonic
generation signal of the drive that near resonance with the signal
mode.}
\label{Fig1}
\end{figure}

In this Letter, we propose a novel method of achieving resonant-enhanced
optical frequency conversion on a photonic chip via degenerate sum-frequency
process using a single input mode. Since both drive and signal are
near resonance with the same telecom mode, only the two-mode phase
matching condition is required in this conversion which is flexible
for experiments. Through this scheme, high photon-number conversion
efficiency of $42\%$ is realized, with a full-width-at-half-maximum
(FWHM) bandwidth of the tuning range over $250\,\text{GHz}$. Furthermore,
in the high-power regime, we reveal that the converted signal is amplified
due to the cascaded nonlinear optical effects in the same device.
Our results of the cross-band frequency conversion and amplification
in a single microresonator manifest the advantages of on-chip nonlinear
photonic devices in both fundamental researches and potential applications.

\emph{Principle.-} Within a microring resonator composed of materials
with $\chi^{(2)}$ nonlinearity, there are many modes in both visible
and IR frequency bands~\citep{Guo2016,ChenLN2019}. In general, the
nonlinear coupling between the modes due to the $\chi^{(2)}$ nonlinearity
can be described by the interaction Hamiltonian~\citep{Guo2016,Breunig2016}
\begin{equation}
H_{\chi^{(2)}}=\sum_{j,k,l}g_{jkl}\left(a_{j}a_{k}b_{l}^{\dagger}+h.c.\right),
\end{equation}
where $a_{j}$, $b_{k}$ denote the bosonic operators of the optical
modes, with subscripts $j,k\in\mathbb{Z}$ as the mode orbit angular
momentum, and $g_{jkl}$ represents the coupling strength. According
to the momentum conservation~\citep{Guo2016a}, $g_{jkl}$ is nontrivial
only for $j+k=l$. For the telecom and visible frequency ranges we
are interested in, the mode profiles are similar, thus $g_{jk(j+k)}$
is approximated by a constant $g$ in the following. For non-degenerate
sum-frequency generation (SFG), as shown in Fig.$\,$\ref{Fig1},
we drive one IR mode with an intracavity amplitude $\alpha_{j}$,
then we realize the frequency conversion by the linearized Hamiltonian~\citet{Kumar1990}
$H_{\mathrm{FC}}=\sum_{k}\left(G_{j}a_{k}b_{j+k}^{\dagger}+G_{j}^{*}a_{k}^{\dagger}b_{j+k}\right)$,
which induces coherent coupling between $a_{k}$ and $b_{l}$ of distinct
wavelengths, while $G_{j}=2g_{2}\alpha_{j}$ is the stimulated coupling
strength. However, the non-degenerate SFG requires a triply-resonant
condition ($\omega_{a_{j}}+\omega_{a_{k}}\approx\omega_{b_{j+k}}$)
for energy conservation. In practical applications for converting
the telecom photons to match the frequency of narrow atomic transition
at $\omega_{b_{l}}$, the stringent triply-resonant condition imposes
significant challenges for finding modes that satisfy all requirements.
\begin{figure*}[t]
\begin{centering}
\includegraphics[width=1\textwidth]{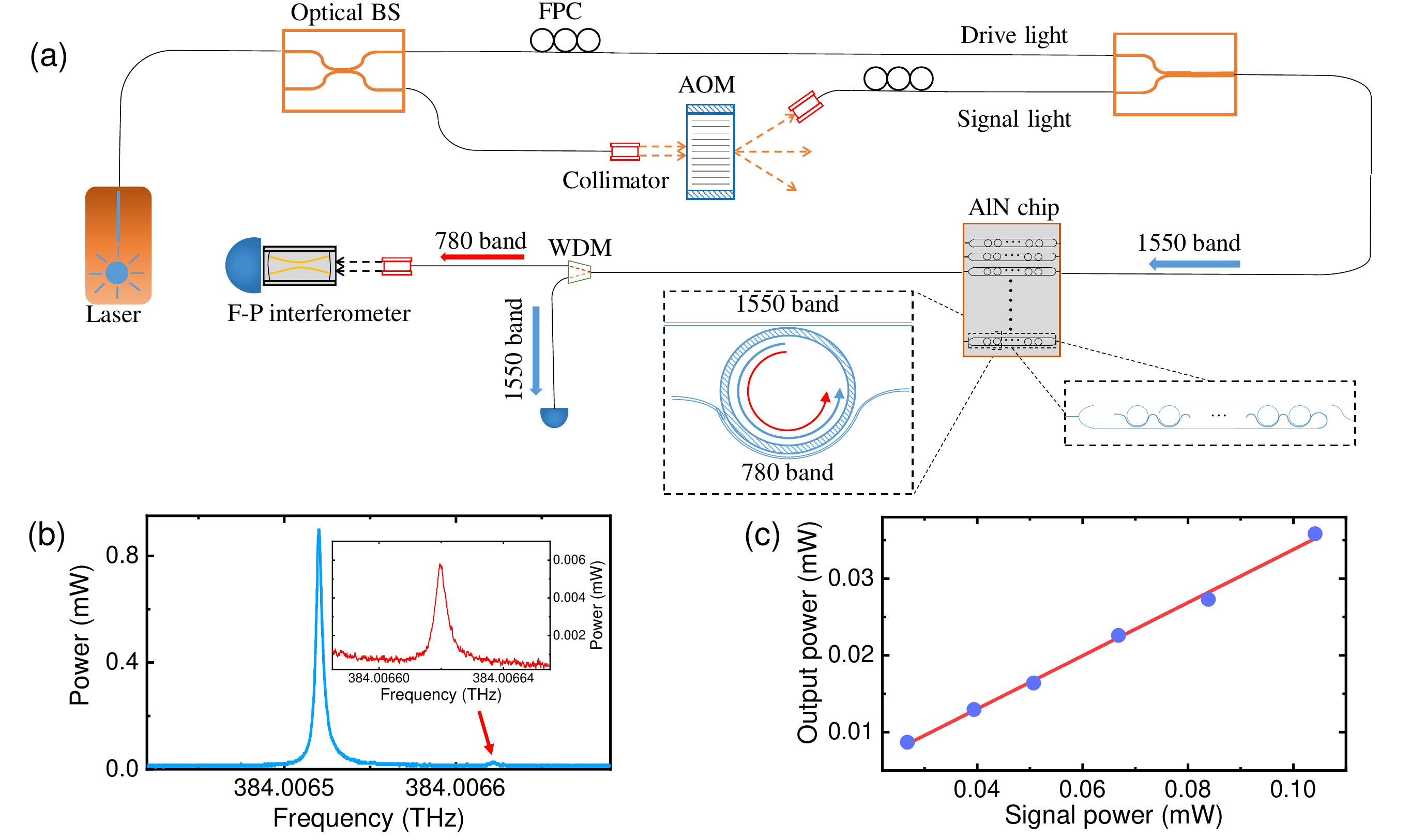}
\par\end{centering}
\caption{(a) Experimental setup. PD, photodetector; Optical BS, optical beam
splitter; FPC, fiber polarization controller; AOM, acoustic-optical
modulator; WDM, wave division multiplexing. (b) Visible output spectra
for degenerate SFG. The blue represents the results by the detector
with low gain. The strong peak represents the SHG of drive light,
and the right small peak corresponds to the converted signal. Inset:
the converted signal probed by the detector with high gain. (c) The
relationship between input and output signal on-chip powers. The blue
dots are experimental values, and the red line is a linear fit.}

\label{Fig2}
\end{figure*}

Instead of demanding matching conditions for non-degenerate modes
in SFG, its degenerate counterpart, which was known as second-harmonic
generation (SHG), relaxes the requirements on phase matching and energy
matching ($2j=l$ and $2\omega_{a_{j}}\approx\omega_{b_{l}}$) because
fewer optical modes participate in the interaction. However, SHG is
not a linear process, so the conversion is only efficient when signal
power is high enough. Therefore, we propose to drive the mode $a_{j}$
near-resonance and couple the on-resonance signal to the same mode,
thus realizing the degenerate SFG (Fig.$\,$\ref{Fig1}) with the
Hamiltonian as 
\begin{equation}
H_{\mathrm{FC}}=\sum_{j}\left(Ga_{j}b_{2j}^{\dagger}+G^{*}a_{j}^{\dagger}b_{2j}\right),\label{eq:FC}
\end{equation}
which indicates a linear signal conversion process.

\emph{Experimental setup.-} To demonstrate the frequency conversion
by the new scheme, we carried out experiments by an AlN photonic chip~\citep{Guo2016,Guo2017,Surya2018,Bruch2019},
with the experimental setup shown in Fig.$\,$\ref{Fig2}(a). The
microrings on the chip are optimized for SHG, with the phase-matching
conditions satisfying telecom wavelengths~\citep{Surya2018}. A telecom
laser is amplified by an erbium-doped fiber amplifier and divided
into two beams for drive and signal, respectively. The signal is generated
from the laser through an acoustic-optical modulator (AOM, $\sim100$~MHz)
with the frequency shifted from the drive laser, and is collected
by a fiber, since the sideband of the AOM is deflected from the input.
Then, both drive and signal light are combined through a beam splitter
and seeded to the AlN chip through a fiber lens. The converted signal
and the generated SHG light (around $780\,\mathrm{nm}$), as well
as transmitted drive (around $1560\,\mathrm{nm}$), are collected
by fiber lens on the output end of the chip and separated by a wavelength
division multiplexer (WDM), and finally collected by photodetectors.
Prior to the frequency conversion experiments, we identified an appropriate
device to achieve high SHG efficiency by sweeping the drive.

To validate the frequency conversion via degenerate SFG, we send both
drive and signal to the selected mode and measured the converted signal
at visible wavelength, with the visible spectrum measured by a scanning
Fabry--P�rot (F-P) interferometer. As shown in Fig.$\,$\ref{Fig2}(b),
on the background of a strong SHG peak due to the drive, there is
a small peak of about $100\,\mathrm{MHz}$ detuned, corresponding
to $22.4\,\mathrm{mW}$ and $0.096\,\mathrm{mW}$ for the calibrated
on-chip power of drive and signal, respectively. Furthermore, to confirm
that the weak peak is caused by the SFG rather than other effects,
we explore the relationship between the power of the weak peak ($P_{\mathrm{Vis}}$)
at visible frequency and the power of input IR signal ($P_{\mathrm{IR}}$).
The experimental results in Fig.~\ref{Fig2}(c) are well fitted by
a linear function $P_{\mathrm{Vis}}=0.346P_{\mathrm{IR}}-7.9\times10^{-4}\,\mathrm{mW}$,
in contrast to the $P_{\mathrm{IR}}^{2}$-scaling for SHG, with a
slope $0.346$ is the power efficiency and a negligible residual due
to the noise of detector and strong SHG background. The linear relationship
confirms the single-mode frequency conversion mechanism.
\begin{figure}[h]
\begin{centering}
\includegraphics[width=1\columnwidth]{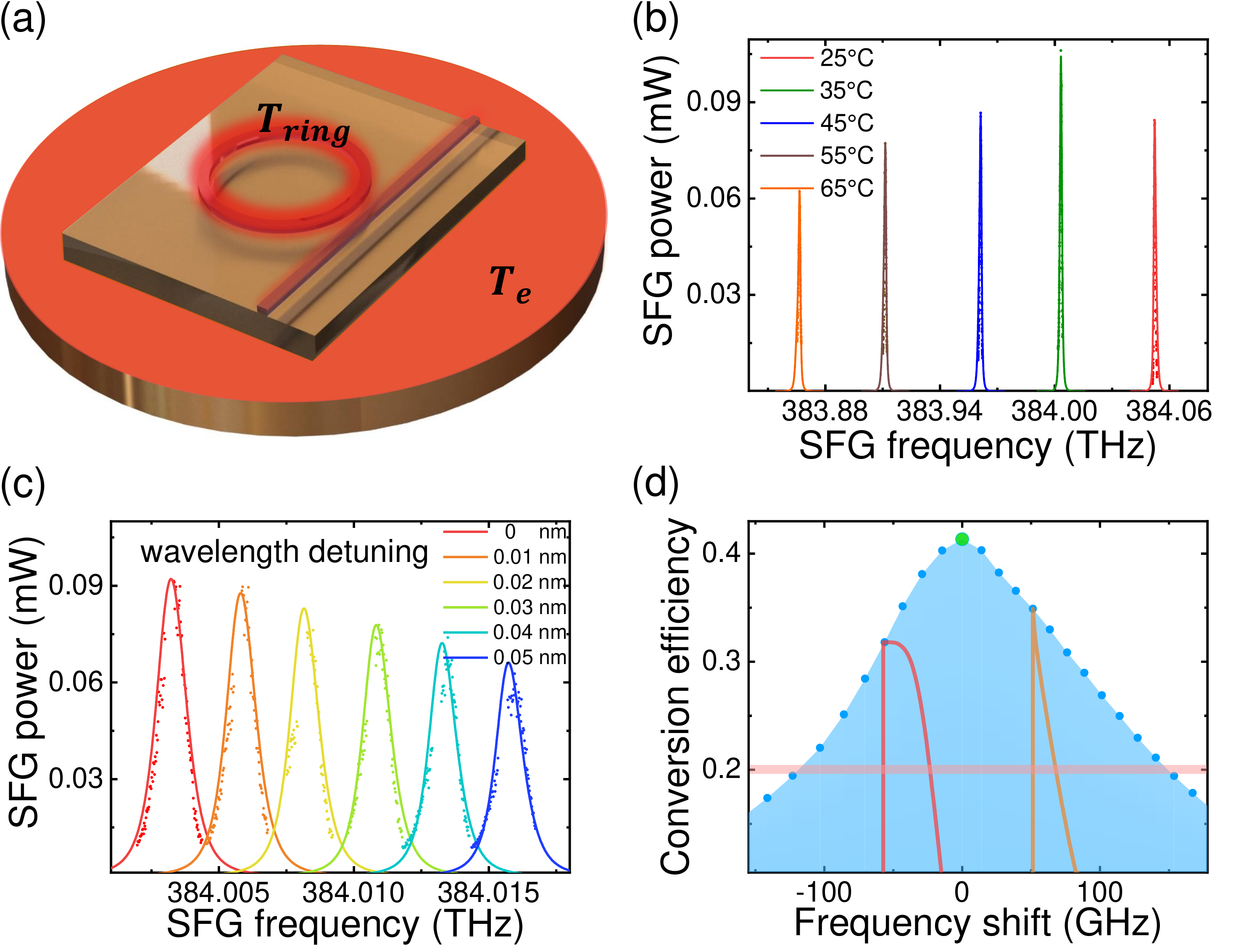}
\par\end{centering}
\caption{Frequency tuning. (a) The schematic of the temperature controlling
for the on-chip microring resonators. The environment temperature
$(T_{\mathrm{e}})$ could be controlled by a heater, and the device
temperature $(T_{\mathrm{ring}})$ could also be adjusted by drive
laser. (b) The coarse tuning of frequency conversion by $T_{\mathrm{e}}$,
with the on-chip input drive and signal powers are about $23\,\mathrm{mW}$
and $0.28\,\mathrm{mW}$, respectively. (c) Fine tuning via $T_{\mathrm{ring}}$
by changing the drive wavelength. In both (b) and (c), the dots are
the experimental data, and the lines are the calculated results based
on device parameters. (d) The prediction of SFG frequency conversion
efficiency for different $T_{\mathrm{e}}$ and drive detuning. The
zero frequency detuning corresponding to the optimal conversion ($\sim40\%$),
and the input on-chip drive power is $30\text{\,\ensuremath{\mathrm{mW}}}$.
The pink line presents $20\%$ efficiency, which indicates a full-width-at-half-maximum
(FWHM) of $257\,\mathrm{GHz}$. The red and yellow lines represent
the typical results for fine tuning under different $T_{\mathrm{e}}$
($16\,\text{K}$ and $-16\,\text{K}$ away from optimal temperature,
respectively).}

\label{Fig3}
\end{figure}

\emph{Precise frequency tuning.- }In a microring resonator, the small
mode volume is beneficial for nonlinear optical frequency conversion,
while it also leads to sparse resonances that impose challenges in
practical applications: the resonant frequency determined by cavity
geometry is usually different from the wavelength we need, i.e. the
target wavelength can not match the frequency window (FW) of the device
for high-efficiency conversion. Therefore, for a useful interface
that connects different wavebands, the precise tuning of FW is demanded.
The temperature dependence of the mode resonant wavelength offers
a convenient way to resolve this problem. Because of the different
mode dispersion at visible ($\omega_{b}$) and telecom ($\omega_{a}$)
band, the resonant frequencies of two modes will shift at different
rates when changing the device temperature, with the expression $\omega_{a(b)}(T_{\mathrm{ring}})=\omega_{a(b)}(T_{\mathrm{e}})+K_{a(b)}(T_{\mathrm{ring}}-T_{\mathrm{e}})$.
Here, $K_{a(b)}$ is the thermal drift coefficient of mode $a\,(b)$,
$T_{\mathrm{ring}}$ is the effective temperature of the microingresonator,
and $T_{\mathrm{e}}$ is the environment temperature of the chip which
could be adjusted by a heater in our experiment, as shown in Fig.$\,$\ref{Fig3}(a).
The relative temperature difference $\delta T=T_{\mathrm{ring}}-T_{\mathrm{e}}\simeq c_{a}|\alpha|^{2}$
is induced by the opto-thermal effect, in which $\alpha$ represents
intracavity photon amplitude and $c_{a}$ is temperature coefficient
of mode $a$ ~\citep{Hu2020}. As a result, there are two approaches
to regulate FW: coarse tuning by changing $T_{\mathrm{e}}$ and keeping
$\delta T$ constant, or fine tuning by the drive light to change
$\delta T$ and output frequency directly. So, in experiments, we
fixed the drive power as a constant and regulated SFG frequency by
changing $T_{\mathrm{e}}$ and the intracavity drive light to characterize
the FW tuning ability.

As shown in Fig.$\,$\ref{Fig3}(b), the SFG signals are generated
at different frequencies within about one cavity linewidth FW by varying
$T_{\mathrm{e}}$ while keeping the drive near-resonance. As expected,
the central frequency of the FW decreases with the increasing temperature,
with a tuning range of FW over $150\,\mathrm{GHz}$ by changing $T_{\mathrm{e}}$
from $25\,\mathrm{^{\circ}C}$ to $65\,\mathrm{^{\circ}C}$. Noting
that the conversion efficiency also changes with temperature, but
the peak conversion efficiency maintains in this large frequency tuning
range. Due to the large chip volume, the precise tuning of the $T_{\mathrm{e}}$
is a challenge and also the thermal response of the whole system is
very slow. Therefore, we adopt the other approach for fine tuning
the SFG signal frequency. As shown in Fig.$\,$\ref{Fig3}(c), by
changing the drive laser wavelength, the temperature difference $\delta T$
could be adjusted precisely by the thermal effects of the microring
resonator. Here, the intracavity drive power reduces with detuning,
consequently, the FW shows a blue-shift as the $\delta T$ reduces
and the efficiency also reduces. Comparing these two methods, the
first approach allows a large dynamic range of frequency tuning while
the other approach is suitable for quick and precise controlling.

In our device, the thermal coefficients $K_{a}$ and $K_{b}$ determine
how fast the resonant frequencies shift with temperature, and their
difference $\delta K=K_{a}-K_{b}$ determines the sensitivity of phase-matching
to temperature. To simulate the full frequency conversion bandwidth,
we measure the highest frequency conversion efficiency at $30\thinspace\mathrm{mW}$
and also fit the thermal coefficients as $K_{a}=-1.8\times10^{9}\,\mathrm{Hz/K}$
and $K_{b}=-3.53\times10^{9}\,\mathrm{Hz/K}$. As shown in Fig.$\,$\ref{Fig3}(d),
the green dot represents the highest SFG efficiency and we set its
frequency shift as zero. The shadow predicts the conversion efficiency
we can achieve by changing $T_{\mathrm{e}}$ and shows a FWHM of $257\,\mathrm{GHz}$.
At each $T_{\mathrm{e}}$, the conversion FW and efficiency could
be adjusted by tuning the drive frequency, as shown by the yellow
and red lines in Fig.$\,$\ref{Fig3}(d). Here, because of the thermal
bistability effect~\citep{Hu2020}, the fining tuning shows a sudden
jump of efficiency when the system achieves the highest efficiency,
and the fine tuning bandwidth varies with the temperature $T_{\mathrm{e}}$.

\begin{figure}
\begin{centering}
\includegraphics[width=1\columnwidth]{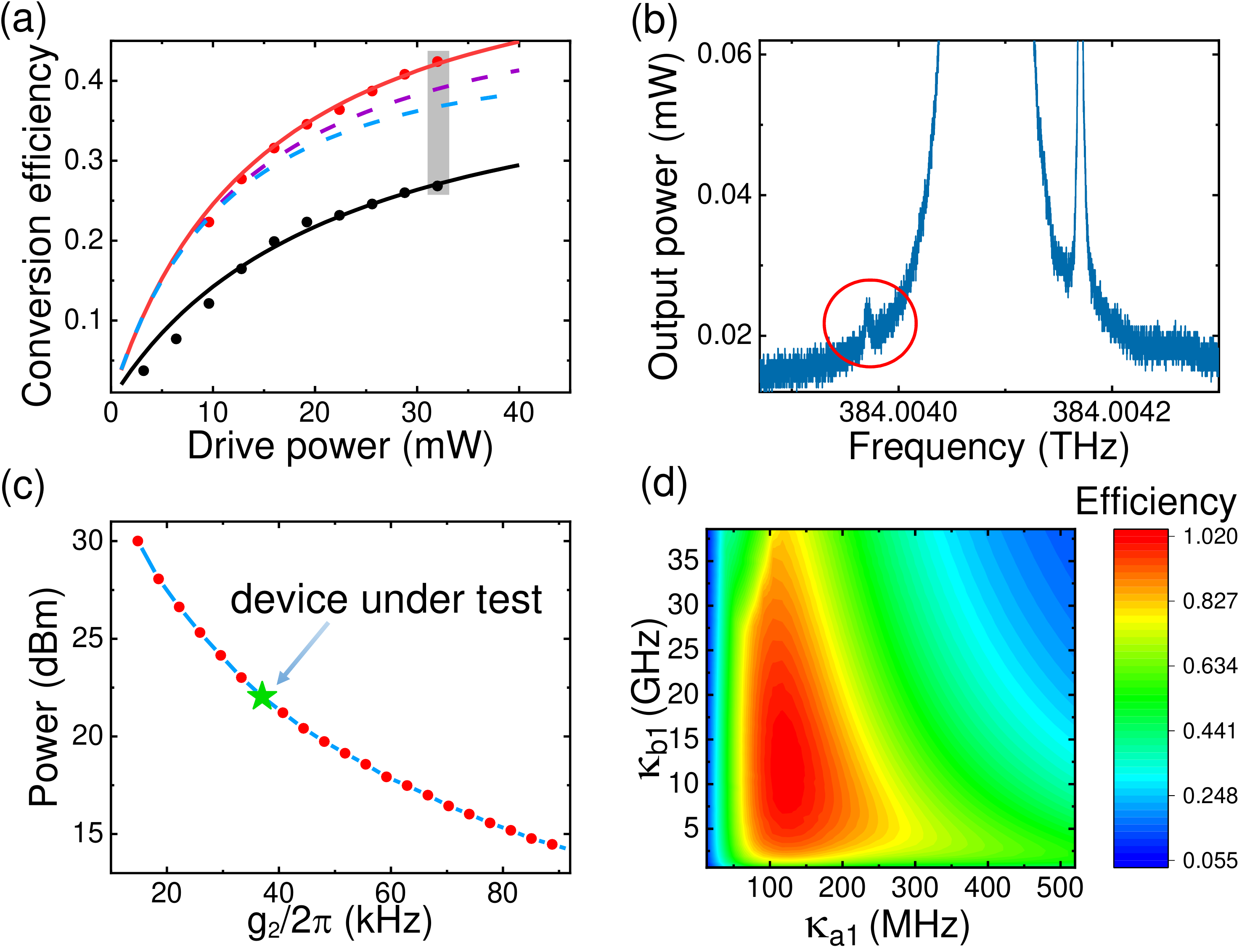}
\par\end{centering}
\caption{Cascaded nonlinear effect in frequency conversion. (a) The conversion
efficiency of SHG and SFG. The dots are experimental data, and the
lines are theoretical predictions. The black line represents the SHG
efficiency, the blue dash line is based on the SFG model without cascaded
nonlinear effect modification, the purple dash line represents the
prediction by including the cascaded $\chi^{(2)}-\chi^{(2)}$ process,
and the red solid line represents the theoretical results for the
full model that including both $\chi^{(2)}-\chi^{(2)}$ and $\chi^{(3)}$
effects. (b) The detailed transmission spectrum at infrared band for
degenerate SFG with drive power of $32\,\mathrm{mW}$, which corresponds
to the shadow region in (a). (c) The prediction of the required drive
power for achieving $100\%$ conversion efficiency for different $g_{2}$,
with $g_{3}$ varies correspondingly and all other parameters are
from the actual device.(d) The parameter dependence between the conversion
efficiency and the external coupling rates $\kappa_{a1},\kappa_{b1}$,
with all other parameters from the actual device and the drive power
is $50\,\mathrm{mW}$.}

\label{Fig4}
\end{figure}

\emph{Cascaded $\chi^{(2)}$ and $\chi^{(3)}$ in frequency conversion.-}
According to the physics presented in Fig.$\,$\ref{Fig1}, the efficiency
of the frequency conversion could be enhanced by the drive laser.
In Fig.$\,$\ref{Fig4}(a), we tested the on-chip conversion efficiency
against the drive power at optimal working condition. The black and
red dots representing the SHG and SFG conversion efficiencies, both
increase with the drive power and show saturation effects when drive
power ($P_{\mathrm{d}}$) exceeding $20\,\mathrm{mW}$. However, when
fitting the experimental results, the SHG agrees with our theoretical
prediction (black curve), while the SFG can not be fitted (blue curve).
It is surprising that the best achieved conversion efficiency $42\%$
is about $15\%$ higher than the prediction (blue curve), which indicates
an unexpected physical process that amplified the signal.

These experimental results could be explained by the cascaded $\chi^{(2)}$
and $\chi^{(3)}$ effects which indicate the possible multiple nonlinear
optics process in a single microresonator. In the microresonator,
not only the $\chi^{(2)}$ is enhanced for both SHG and down-conversion
processes~\citep{Guo2016,Guo2016a,Guo2017}, but also the $\chi^{(3)}$
is enhanced as the phasing matching condition naturally satisfied
for single mode ($j+j=j+j$) . When the drive and signal are injected
into in the same mode, the cascaded SHG and down-conversion ($\chi^{(2)}-\chi^{(2)}$)
processes contribute an effective $\chi^{(3)}$-process, which can
not be neglected when the conversion efficiency is high~\citep{Li2018}.
So, there is a modification of the Hamiltonian to degenerate SFG as
\begin{eqnarray}
H_{\mathrm{Cascade}} & = & \left(G_{a}a^{\dagger2}+G_{a}^{*}a^{2}\right)+H_{\mathrm{FC}},\label{eq:Effective}
\end{eqnarray}
where $G_{a}=g_{2}\beta+2g_{3}\alpha^{2}$, and $g_{3}$ is the vacuum
coupling strength for the self-Kerr effect~\citep{supplement}. The
$g_{2}\beta$ corresponds to the cascaded $\chi^{(2)}-\chi^{(2)}$
effect with $\beta$ being the SHG field of drive laser and $2g_{3}\alpha^{2}$
represents the intrinsic $\chi^{(3)}$ effect. Therefore, the cascaded
effect of the effective $\chi^{(3)}$ process and the degenerate SFG
would induce an amplification of signal when realizing the frequency
conversion, and the amplification is non-negligible for high drive
power ($G_{a}/G_{b}=\mathcal{O}\left(1\right)$). By including these
cascaded nonlinear effects, the experimental results are explained
well by the red curve in Fig.$\,$\ref{Fig4}(a). To further verify
the cascaded nonlinear optical processes, we measured the output spectrum
by the F-P interferometer with $P_{\mathrm{d}}=38\,\mathrm{mW}$.
As expected, we find an additional signal peak on the left-side of
the drive SHG light, with the frequency being symmetric with SFG light,
as shown in the red circle in Fig.$\,$\ref{Fig4}(b). Here, the on-chip
drive power and conversion efficiency of Fig.$\,$\ref{Fig4}(b) are
corresponding to the shadow part of Fig.$\,$\ref{Fig4}(a).

Due to the power limitation in our setup, the amplification effect
is not strong enough and only generates a relatively small peak. To
further explore the influence of the cascaded nonlinear effect in
our experiment, we simulate the parameter dependence of conversion
efficiency. As shown in Fig.$\,$\ref{Fig4}(c), we scale $g_{2}\propto1/\sqrt{R}$
and $g_{3}\propto1/R$ with the $R$ representing radius of microring~\citep{Guo2018kerr}
to study their relationship with conversion efficiency. With the $g_{2,3}$
increasing, the power required to achieve $100\%$ conversion efficiency
decreases rapidly. The green pentacle in Fig.$\,$\ref{Fig4}(c) indicates
the parameters of our actual device. Besides, we studied the dependence
of conversion efficiency on the external coupling rates $\kappa_{a1},\kappa_{b1}$,
while keeping the intrinsic loss rate $\kappa_{a0},\,\kappa_{b0}$
fixed, because the intrinsic loss rates are mainly determined by the
material itself and the fabrication technique while the external coupling
rates could be adjusted by changing the device geometry. The results
are plotted in Fig.$\,$\ref{Fig4}(d), which show the highest conversion
efficiency over $100\%$. These results indicate that by optimizing
external coupling rates and intrinsic nonlinear coupling strengths,
we can realize high frequency conversion efficiency for cross-band
signal amplification.

\emph{Conclusion.-} Based on aluminum nitride microring, we have achieved
high-efficiency frequency conversion via degenerate sum-frequency
process. The conversion efficiency is up to $42\%$, with a frequency
tuning bandwidth over $250\,\mathrm{GHz}$ for efficiency exceeding
$20\%$. The demonstrated degenerate SFG scheme is feasible for a
wide range of devices and materials, thus allows the realization of
high-efficiency frequency conversion for future quantum communications,
atom clocks, and sensing applications. Further, we demonstrate the
cascaded nonlinear effect and the induced amplification of the converted
signal in the same microring resonator, which reveal the advantage
of integrated chip in frequency conversion and deserve further investigation
from both applied and fundamental points of view.\smallskip{}

\noindent \textbf{Acknowledgments}\\This work was funded by the National
Key Research and Development Program (Grant No. 2016YFA0301300) and
the National Natural Science Foundation of China (Grant No.11874342,
11934012, 11904316, 11947234, and 11922411), and Anhui Initiative
in Quantum Information Technologies (AHY130200). CLZ was also supported
by the Fundamental Research Funds for the Central Universities, and
the State Key Laboratory of Advanced Optical Communication Systems
and Networks, Shanghai Jiao Tong University, China. This work was
partially carried out at the USTC Center for Micro and Nanoscale Research
and Fabrication.

\clearpage{}

\onecolumngrid
\renewcommand{\thefigure}{S\arabic{figure}}
\setcounter{figure}{0} 
\renewcommand{\thepage}{S\arabic{page}}
\setcounter{page}{1} 
\renewcommand{\theequation}{S.\arabic{equation}}
\setcounter{equation}{0} 
\setcounter{section}{0}

\begin{center}
\textbf{\textsc{\LARGE{}Supplementary Information}}{\LARGE\par}
\par\end{center}

\tableofcontents{}

\section{Theoretical Derivation}

In our experiment, in a $\chi^{(2)}$ microresonator, with one strong
driving light ($\omega_{\mathrm{d}}$) and one weak signal light ($\omega_{\mathrm{s}}$)
coupling in one mode, the Hamiltonian reads 
\begin{eqnarray}
H & = & \omega_{a}a^{\dagger}a+\omega_{b}b^{\dagger}b+g_{2}\left(a^{2}b^{\dagger}+a^{\dagger2}b\right)+g_{3}(a^{\dagger}a^{\dagger}aa+aaa^{\dagger}a^{\dagger})\nonumber \\
 &  & +i\varepsilon_{\mathrm{d}}\left(a^{\dagger}e^{-i\omega_{\mathrm{d}}t}+ae^{i\omega_{\mathrm{d}}t}\right)+i\varepsilon_{\mathrm{s}}\left(a^{\dagger}e^{-i\omega_{\mathrm{s}}t}+a{}^{i\omega_{\mathrm{s}}t}\right)
\end{eqnarray}
where $\varepsilon_{\mathrm{d}}=\sqrt{2\kappa_{a,1}P_{\mathrm{d}}/\hbar\omega_{\mathrm{d}}}$
and $\varepsilon_{\mathrm{s}}=\sqrt{2\kappa_{a,1}P_{\mathrm{s}}/\hbar\omega_{\mathrm{s}}}$
denotes the drive and probe field amplitude, respectively, and $P_{\mathrm{d}}$($P_{\mathrm{s}}$)
is the drive (signal) power. $g_{2}$ and $g_{3}$ are the nonlinear
coupling strength due to the second- and third-order nonlinearity,
respectively. $\kappa_{a}=\kappa_{a0}+\kappa_{a1}$ ($\kappa_{b}=\kappa_{b0}+\kappa_{b1}$)
is the total amplitude loss rate of cavity mode $a(b)$, including
the external coupling rate $\kappa_{a(b)1}$ and the intrinsic loss
rate $\kappa_{a(b)0}$.

In the rotating frame of $\omega_{\mathrm{d}}a^{\dagger}a+2\omega_{\mathrm{d}}b^{\dagger}b$,
the Hamiltonian reduces to
\begin{eqnarray}
H & = & \Delta_{a}a^{\dagger}a+\Delta_{b}b^{\dagger}b+g_{2}\left(a^{2}b^{\dagger}+a^{\dagger2}b\right)+g_{3}(a^{\dagger}a^{\dagger}aa+aaa^{\dagger}a^{\dagger})+i\varepsilon_{\mathrm{d}}\left(a^{\dagger}+a\right)+i\varepsilon_{\mathrm{s}}\left(a^{\dagger}e^{-i\delta_{\mathrm{s}}t}+ae^{i\delta_{\mathrm{s}}t}\right)
\end{eqnarray}
where $\Delta_{a}=\omega_{a}-\omega_{\mathrm{d}}$, $\Delta_{b}=\omega_{b}-2\omega_{\mathrm{d}}$,
and $\delta_{\mathrm{s}}=\omega_{\mathrm{s}}-\omega_{\mathrm{d}}$.

First of all, we make the first order approximation of the system
by neglecting the weak field term $\varepsilon_{\mathrm{s}}$, since
$|\varepsilon_{\mathrm{d}}|\gg|\varepsilon_{\mathrm{s}}|$. Then,
we can apply the mean filed approximation: $a\rightarrow\alpha+a,\,b\rightarrow\beta+b$,
and the mean fields satisfy the equations
\begin{eqnarray}
\frac{d}{dt}\alpha & = & \left(-i\Delta_{a}-\kappa_{a}\right)\alpha-2ig_{2}\alpha^{*}\beta-4ig_{3}\alpha^{*}\alpha^{2}-i\varepsilon_{\mathrm{d}},\\
\frac{d}{dt}\beta & = & \left(-i\Delta_{b}-\kappa_{b}\right)\beta-ig_{2}\alpha^{2}.
\end{eqnarray}
At steady state, $\alpha$ and $\beta$ can be solved as 
\begin{align}
\alpha & =\frac{i\varepsilon_{\mathrm{d}}}{-i(\Delta_{a}+4g_{3}|\ensuremath{\alpha|^{2}})-\kappa_{a}+\frac{2g_{2}^{2}|\ensuremath{\alpha|^{2}}}{-i\Delta_{b}-\kappa_{b}}},\\
\beta & =\frac{ig_{2}}{-i\Delta_{b}-\kappa_{b}}\alpha^{2}.
\end{align}

For the weak fluctuations, we arrive at the effective Hamiltonian
\begin{eqnarray}
H_{\mathrm{eff}} & = & \Delta_{a}a^{\dagger}a+\Delta_{b}b^{\dagger}b+\left((g_{2}\beta+2g_{3}\alpha^{2})a^{\dagger2}+(g_{2}\beta^{*}+2g_{3}\alpha^{*2})a^{2}\right)+2g_{2}\left(\alpha^{*}a^{\dagger}b+\alpha ab^{\dagger}\right)\nonumber \\
 & = & \Delta_{a}a^{\dagger}a+\Delta_{b}b^{\dagger}b+\left(G_{a}a^{\dagger2}+G_{a}^{*}a^{2}\right)+\left(G_{b}ab^{\dagger}+G_{b}^{*}a^{\dagger}b\right),\label{eq:hamiltonian}
\end{eqnarray}
with the stimulated coupling strengths 
\begin{align}
G_{a} & =g_{2}\beta+2g_{3}\alpha^{2},\\
 & =\alpha^{2}(\frac{ig_{2}^{2}}{-i\Delta_{b}-\kappa_{b}}+2g_{3}),\label{eq:Ga-expression}
\end{align}
and $G_{b}=2g_{2}\alpha$, the input $a_{\mathrm{in}}=a_{\mathrm{\mathrm{in},0}}+a_{\mathrm{in,1}}$,
where $a_{\mathrm{in,1}}$ and $a_{\mathrm{in,0}}$ represents the
coherent weak signal and vacuum noise, respectively. The dynamics
of weak signal fields follow the equations
\begin{eqnarray}
\frac{d}{dt}a & = & \left(-i(\Delta_{a}+8g_{3}|\alpha|^{2})-\kappa_{a}\right)a-2ig_{2}\beta a^{\dagger}-4ig_{3}\alpha^{2}a^{\dagger}-2ig\alpha^{*}b+a_{\mathrm{in}},\\
\frac{d}{dt}b & = & \left(-i\Delta_{b}-\kappa_{b}\right)b-2ig\alpha a+b_{\mathrm{in}}.
\end{eqnarray}
For simplicity, we denote $(\Delta_{a}+8g_{3}|\alpha|^{2})$ by $\Delta_{a,\mathrm{eff}}$,
and we have
\begin{eqnarray}
\frac{d}{dt}a & = & \left(-i\Delta_{a,\mathrm{eff}}-\kappa_{a}\right)a-2iG_{a}a^{\dagger}-iG_{b}^{*}b+A_{\mathrm{in}},\\
\frac{d}{dt}a^{\dagger} & = & \left(i\Delta_{a,\mathrm{eff}}-\kappa_{a}\right)a^{\dagger}+2iG_{a}^{*}a+iG_{b}b^{\dagger}+A_{\mathrm{in}}^{\dagger},\\
\frac{d}{dt}b & = & \left(-i\Delta_{b}-\kappa_{b}\right)b-iG_{b}a+B_{\mathrm{in}},\\
\frac{d}{dt}b^{\dagger} & = & \left(i\Delta_{b}-\kappa_{b}\right)b^{\dagger}+iG_{b}^{*}a^{\dagger}+B_{\mathrm{in}}^{\dagger},
\end{eqnarray}
with $A_{\mathrm{in}}=\sqrt{2\kappa_{a,1}}a_{\mathrm{in},1}+\sqrt{2\kappa_{a,0}}a_{\mathrm{in},0}$
and $B_{\mathrm{in}}=\sqrt{2\kappa_{b}}b_{\mathrm{in}}$. Performing
the Fourier transformation
\begin{eqnarray}
a(\omega) & = & \frac{1}{2\pi}\int dte^{-i\omega t}a(t),\\
a^{\dagger}(-\omega) & = & \frac{1}{2\pi}\int dte^{-i\omega t}a^{\dagger}(t),
\end{eqnarray}
we get the coupled equation between $a(\omega)$, $b(\omega)$ and
$a^{\dagger}(-\omega)$, $b^{\dagger}(-\omega)$, in a compact form
\begin{equation}
\mathbf{M}\cdot\mathbf{A}+\mathbf{A}_{\mathrm{in}}=0,
\end{equation}
with 
\begin{equation}
\mathbf{A}=\left(\begin{array}{c}
a\left(\omega\right)\\
a^{\dagger}\left(-\omega\right)\\
b\left(\omega\right)\\
b^{\dagger}\left(-\omega\right)
\end{array}\right),\;\mathbf{A}_{\mathrm{in}}=\left(\begin{array}{c}
A_{\mathrm{in}}\left(\omega\right)\\
A_{\mathrm{in}}^{\dagger}\left(-\omega\right)\\
B_{\mathrm{in}}\left(\omega\right)\\
B_{\mathrm{in}}^{\dagger}\left(-\omega\right)
\end{array}\right),
\end{equation}
and
\begin{equation}
\mathbf{M}=\left(\begin{array}{cccc}
-i\left(\Delta_{a,\mathrm{eff}}+\omega\right)-\kappa_{a} & -2iG_{a} & -iG_{b}^{*} & 0\\
2iG_{a}^{*} & i\left(\Delta_{a,\mathrm{eff}}-\omega\right)-\kappa_{a} & 0 & iG_{b}\\
-iG_{b} & 0 & -i\left(\Delta_{b}+\omega\right)-\kappa_{b} & 0\\
0 & iG_{\mathrm{b}}^{*} & 0 & i\left(\Delta_{b}-\omega\right)-\kappa_{b}
\end{array}\right).
\end{equation}

The equation array has solution of the following form
\begin{eqnarray}
\mathbf{A} & = & \mathbf{F}\left(\omega\right)\cdot\mathbf{A}_{\mathrm{in}}.
\end{eqnarray}
The intracavity field spectral in mode $b$ is 
\begin{eqnarray}
S_{b}\left(\omega,\omega'\right) & = & \langle b^{\dagger}\left(\omega\right)b\left(\omega'\right)\rangle\nonumber \\
 & = & \langle\left(F_{41}\left(-\omega\right)A_{\mathrm{in}}\left(-\omega\right)+F_{42}\left(-\omega\right)A_{\mathrm{in}}^{\dagger}\left(\omega\right)+F_{43}\left(-\omega\right)B_{\mathrm{in}}\left(-\omega\right)+F_{44}\left(-\omega\right)B_{\mathrm{in}}^{\dagger}\left(\omega\right)\right)\cdot\nonumber \\
 &  & \left(F_{31}\left(\omega'\right)A_{\mathrm{in}}\left(\omega'\right)+F_{32}\left(\omega'\right)A_{\mathrm{in}}^{\dagger}\left(-\omega'\right)+F_{33}\left(\omega'\right)B_{\mathrm{in}}\left(\omega'\right)+F_{34}\left(\omega'\right)B_{\mathrm{in}}^{\dagger}\left(-\omega'\right)\right)\rangle\nonumber \\
 & = & F_{41}\left(-\omega\right)F_{32}\left(\omega'\right)\langle A_{\mathrm{in}}\left(-\omega\right)A_{\mathrm{in}}^{\dagger}\left(-\omega'\right)\rangle+F_{42}\left(-\omega\right)F_{31}\left(\omega'\right)\langle A_{\mathrm{in}}^{\dagger}\left(\omega\right)A_{\mathrm{in}}\left(\omega'\right)\rangle\\
 &  & +F_{43}\left(-\omega\right)F_{34}\left(\omega'\right)\langle B_{\mathrm{in}}\left(-\omega\right)B_{\mathrm{in}}^{\dagger}\left(-\omega'\right)\rangle+F_{44}\left(-\omega\right)F_{33}\left(\omega'\right)\langle B_{\mathrm{in}}^{\dagger}\left(\omega\right)B_{i\mathrm{n}}\left(\omega'\right)\rangle\nonumber \\
 &  & +F_{41}\left(-\omega\right)F_{31}\left(\omega'\right)\langle A_{\mathrm{in}}\left(-\omega\right)A_{\mathrm{in}}\left(\omega'\right)\rangle+F_{42}\left(-\omega\right)F_{32}\left(\omega'\right)\langle A_{\mathrm{in}}^{\dagger}\left(\omega\right)A_{\mathrm{in}}^{\dagger}\left(-\omega'\right)\rangle.\nonumber 
\end{eqnarray}
Denote $\chi_{a1}=-i\left(\Delta_{a,\mathrm{eff}}+\omega\right)-\kappa_{a}$,
$\chi_{a2}=i\left(\Delta_{a,\mathrm{eff}}-\omega\right)-\kappa_{a}$,
$\chi_{b1}=-i\left(\Delta_{b}+\omega\right)-\kappa_{b}$, $\chi_{b2}=i\left(\Delta_{b}-\omega\right)-\kappa_{b}$,
it is fulfilled that $\chi_{a1}^{*}\left(-\omega\right)=\chi_{a2}$,
$\chi_{b1}^{*}\left(-\omega\right)=\chi_{b2}$. 
\begin{eqnarray}
F_{41}\left(\omega\right) & = & \frac{2\alpha_{b1}G_{a}^{*}G_{b}^{*}}{|G_{b}|^{4}+\chi_{b1}\chi_{b2}\left(\chi_{a1}\chi_{a2}-4|G_{a}|^{2}\right)+|G_{b}|^{2}\left(\chi_{a1}\chi_{b1}+\chi_{a2}\chi_{b2}\right)},\\
F_{42}\left(\omega\right) & = & \frac{iG_{b}^{*}\left(\alpha_{a1}\alpha_{b1}+|G_{b}|^{2}\right)}{|G_{b}|^{4}+\chi_{b1}\chi_{b2}\left(\chi_{a1}\chi_{a2}-4|G_{a}|^{2}\right)+|G_{b}|^{2}\left(\chi_{a1}\chi_{b1}+\chi_{a2}\chi_{b2}\right)},\\
F_{43}\left(\omega\right) & = & \frac{2iG_{a}^{*}G_{b}^{*2}}{|G_{b}|^{4}+\chi_{b1}\chi_{b2}\left(\chi_{a1}\chi_{a2}-4|G_{a}|^{2}\right)+|G_{b}|^{2}\left(\chi_{a1}\chi_{b1}+\chi_{a2}\chi_{b2}\right)},\\
F_{44}\left(\omega\right) & = & \frac{-\alpha_{a2}|G_{b}|^{2}-\alpha_{b1}\left(\alpha_{a1}\alpha_{a2}-4|G_{a}|^{2}\right)}{|G_{b}|^{4}+\chi_{b1}\chi_{b2}\left(\chi_{a1}\chi_{a2}-4|G_{a}|^{2}\right)+|G_{b}|^{2}\left(\chi_{a1}\chi_{b1}+\chi_{a2}\chi_{b2}\right)},\\
F_{31}\left(\omega\right) & = & F_{42}^{*}\left(-\omega\right),\\
F_{32}\left(\omega\right) & = & F_{41}^{*}\left(-\omega\right),\\
F_{33}\left(\omega\right) & = & F_{44}^{*}\left(-\omega\right),\\
F_{34}\left(\omega\right) & = & F_{43}^{*}\left(-\omega\right).
\end{eqnarray}
The correlation spectral of mode is calculated as
\begin{eqnarray}
S_{b}\left(\omega,\omega'\right) & = & F_{32}^{*}\left(\omega\right)F_{32}\left(\omega'\right)\langle A_{\mathrm{in}}\left(-\omega\right)A_{\mathrm{in}}^{\dagger}\left(-\omega'\right)\rangle+F_{31}^{*}\left(\omega\right)F_{31}\left(\omega'\right)\langle A_{\mathrm{in}}^{\dagger}\left(\omega\right)A_{\mathrm{in}}\left(\omega'\right)\rangle\nonumber \\
 &  & +F_{34}^{*}\left(\omega\right)F_{34}\left(\omega'\right)\langle B_{\mathrm{in}}\left(-\omega\right)B_{\mathrm{in}}^{\dagger}\left(-\omega'\right)\rangle+F_{33}^{*}\left(\omega\right)F_{33}\left(\omega'\right)\langle B_{\mathrm{in}}^{\dagger}\left(\omega\right)B_{\mathrm{in}}\left(\omega'\right)\rangle\nonumber \\
 &  & +F_{32}^{*}\left(\omega\right)F_{31}\left(\omega'\right)\langle A_{\mathrm{in}}\left(-\omega\right)A_{\mathrm{in}}\left(\omega'\right)\rangle+F_{31}^{*}\left(\omega\right)F_{32}\left(\omega'\right)\langle A_{\mathrm{in}}^{\dagger}\left(\omega\right)A_{\mathrm{in}}^{\dagger}\left(-\omega'\right)\rangle.
\end{eqnarray}
In the time domain $\langle a_{\mathrm{in}}\left(t\right)a_{\mathrm{in}}^{\dagger}\left(t'\right)\rangle=\langle a_{\mathrm{in}}^{\dagger}\left(t\right)a_{\mathrm{in}}\left(t'\right)\rangle+\delta\left(t-t'\right)$,
the input terms
\begin{eqnarray}
\langle a_{\mathrm{in},1}^{\dagger}\left(\omega\right)a_{\mathrm{in},1}\left(\omega'\right)\rangle & = & \delta\left(\omega-\Delta_{\mathrm{s}}\right)\delta\left(\Delta_{\mathrm{s}}-\omega'\right)|\varepsilon_{\mathrm{s}}|^{2}/2\kappa_{a,1},\\
\langle a_{\mathrm{in},0}^{\dagger}\left(\omega\right)a_{\mathrm{in},0}\left(\omega'\right)\rangle & = & 0,\\
\langle a_{\mathrm{in}}\left(-\omega\right)a_{\mathrm{in}}^{\dagger}\left(-\omega'\right)\rangle & = & \frac{\delta\left(\omega-\omega'\right)}{2\pi}+\delta\left(-\omega'-\Delta_{\mathrm{s}}\right)\delta\left(\Delta_{\mathrm{s}}+\omega\right)|\varepsilon_{\mathrm{s}}|^{2}/2\kappa_{a,1},\\
\langle a_{\mathrm{in},1}\left(-\omega\right)a_{\mathrm{in,}1}\left(\omega'\right)\rangle & = & \delta\left(\omega'-\Delta_{\mathrm{s}}\right)\delta\left(\Delta_{\mathrm{s}}+\omega\right)\varepsilon_{\mathrm{s}}^{2}/2\kappa_{a,1},\\
\langle a_{\mathrm{in},1}^{\dagger}\left(\omega\right)a_{\mathrm{in},1}^{\dagger}\left(-\omega'\right)\rangle & = & \delta\left(\omega-\Delta_{\mathrm{s}}\right)\delta\left(\Delta_{\mathrm{s}}+\omega'\right)\varepsilon_{\mathrm{s}}^{*2}/2\kappa_{a,1},\\
\langle b_{\mathrm{in}}\left(-\omega\right)b_{\mathrm{in}}^{\dagger}\left(-\omega'\right)\rangle & = & \frac{\delta\left(\omega-\omega'\right)}{2\pi},\\
\langle b_{\mathrm{in}}^{\dagger}\left(\omega\right)b_{\mathrm{in}}\left(\omega'\right)\rangle & = & 0.
\end{eqnarray}
So the power spectrum of the intracavity field in the second-harmonic
mode
\begin{eqnarray}
S\left(\omega\right) & = & S_{b}\left(\omega,\omega\right)\nonumber \\
 & = & \delta\left(\omega+\Delta_{\mathrm{s}}\right)|\varepsilon_{\mathrm{s}}|^{2}F_{32}^{*}\left(\omega\right)F_{32}\left(-\Delta_{\mathrm{s}}\right)+\delta\left(\omega-\Delta_{\mathrm{s}}\right)|\varepsilon_{\mathrm{s}}|^{2}F_{31}^{*}\left(\omega\right)F_{31}\left(\Delta_{\mathrm{s}}\right)\nonumber \\
 &  & +\frac{\delta\left(\omega\right)}{\text{\ensuremath{\pi}}}\left(\kappa_{a}F_{32}^{*}\left(\omega\right)F_{32}\left(\omega\right)+\kappa_{b}F_{34}^{*}\left(\omega\right)F_{34}\left(\omega\right)\right).\label{eq:sw}
\end{eqnarray}
If we only consider the output field near the frequency $\Delta_{\mathrm{s}}$,
only the second term contributes to the output. The output power of
the signal is
\begin{eqnarray}
P_{\mathrm{signal}} & = & 2\kappa_{b,1}\hbar\omega_{b}|\varepsilon_{\mathrm{s}}|^{2}|F_{31}\left(\Delta_{\mathrm{s}}\right)|^{2}\nonumber \\
 & = & 2\kappa_{b,1}\hbar\omega_{b}|\varepsilon_{\mathrm{s}}|^{2}|F_{42}\left(-\Delta_{\mathrm{s}}\right)|^{2}.
\end{eqnarray}

In our experiment, as shown by Fig.$\,$\ref{Fig4}(b) in the main
text, under strong coupling regime, the non-degenerate down-conversion
would generate another sideband of drive , and the output of this
idle power in our model can be expressed as
\begin{eqnarray}
P_{\mathrm{idle}} & = & 2\kappa_{b,1}\hbar\omega_{b}|\varepsilon_{\mathrm{s}}|^{2}|F_{32}\left(-\Delta_{\mathrm{s}}\right)|^{2}\nonumber \\
 & = & 2\kappa_{b,1}\hbar\omega_{b}|\varepsilon_{\mathrm{s}}|^{2}|F_{41}\left(\Delta_{\mathrm{s}}\right)|^{2}.
\end{eqnarray}

\section{Numerical Results}

\subsection{Influence of four-wave mixing}

\begin{figure}[h]
\begin{centering}
\includegraphics[width=0.35\columnwidth]{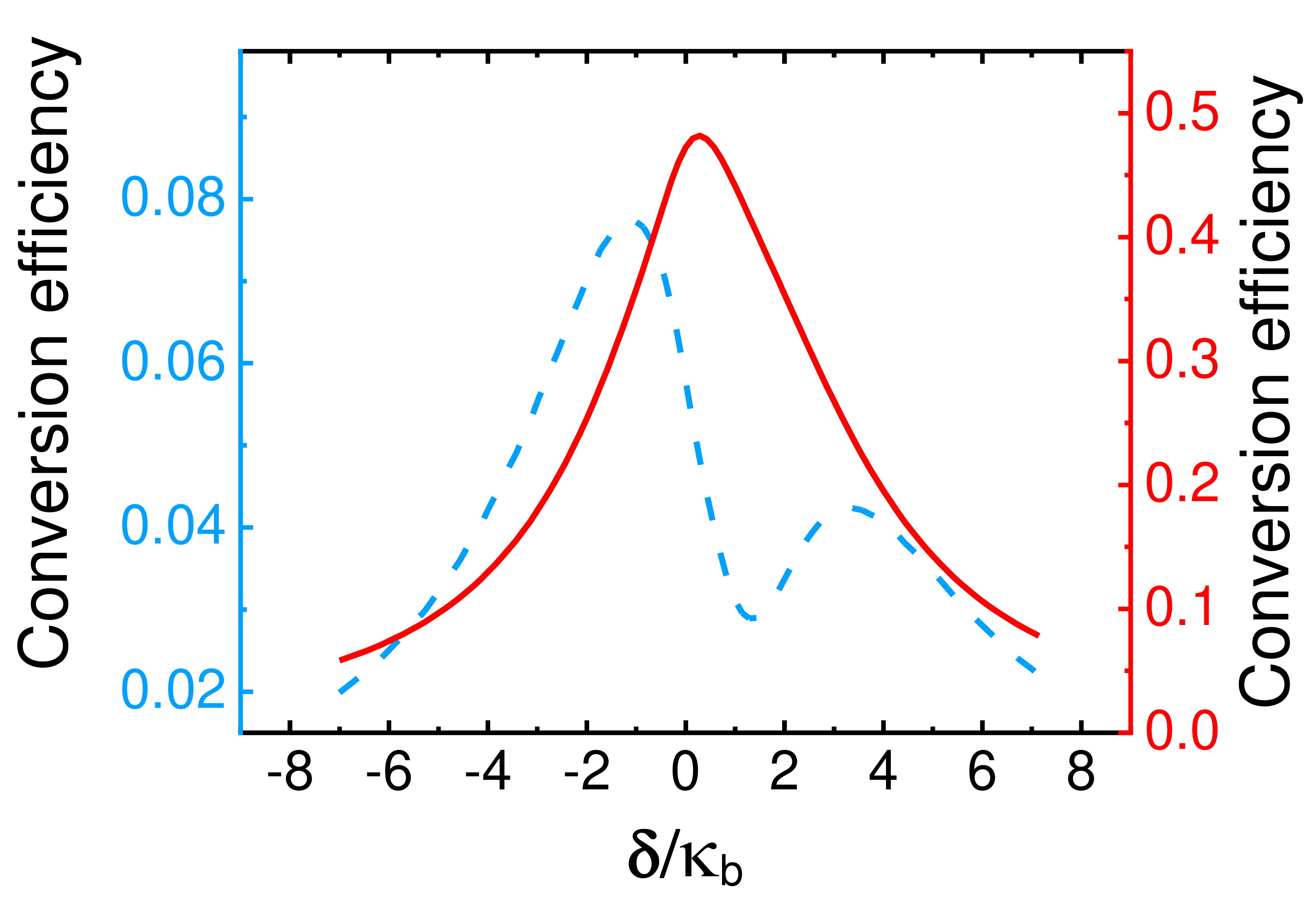}
\par\end{centering}
\caption{The relationship between $\delta$ and conversion efficiency. The
red line represents the total SFG conversion efficiency. The blue
dash line represents the cascaded gain part of conversion corresponding
to $G_{a}$ term. The input on-chip power is 50mW. All of the parameters
are the actual value in our experiment.}

\label{FigS1}
\end{figure}

From the expression of $G_{a}$ {[}Eq.~(\ref{eq:Ga-expression}){]},
there are two processes which induce the parametric gain of photons
in mode $a$, composed by the contributions from the parametric down-conversion
effect by the strong SHG background $\beta$ and the single-mode four-wave
mixing effect (i.e. the self-Kerr effect). The four-wave mixing gives
a coupling strength of $g_{3\mathrm{eff}}=2g_{3}\alpha^{2}$, comparing
with the cascaded $\chi^{(2)}$-process $g_{2}\beta$, we have the
ratio as 
\begin{equation}
\frac{g_{3\mathrm{eff}}}{g_{2}\beta}=\frac{2g_{3}(-i\Delta_{b}-\kappa_{b})}{ig_{2}^{2}}.
\end{equation}
In our experiment, we estimated $g_{2}/2\pi=37\,\mathrm{kHz}$, $\kappa_{b}/2\pi=1.4\,\mathrm{GHz}$,
and $g_{3}/2\pi$ is on the order of Hertz. So, the two contributions
$g_{3\mathrm{eff}}$ and $g_{2}\text{\ensuremath{\beta}}$ are on
the same order of magnitude, thus the influence of the four-wave mixing
effect is not negligible. By this complete model and fitting the experimental
results of SHG conversion efficiency and SFG conversion efficiency,
as shown in Fig.$\,$\ref{Fig4}(a), we obtain the $g_{3}/2\pi\sim0.6\,\mathrm{Hz}$
in our system, which agrees with theoretical expectations. Additionally,
as indicated by Eq.$\,$(\ref{eq:Ga-expression}),
\begin{equation}
G_{a}/\alpha^{2}=\frac{ig_{2}^{2}}{-i\Delta_{b}-\kappa_{b}}+2g_{3},
\end{equation}
the interference between the two effects is feasible for experiments~\citep{Li2018}.
Here, for different detunings $\Delta_{b}$, the two effects would
interfere constructively or destructively. On the other hand, the
$\Delta_{b}$ can be expressed as $\Delta_{b}=\omega_{b}-2\omega_{\mathrm{d}}=\Delta_{a}+(\omega_{b}-2\omega_{a})$.
So in experiment, $\Delta_{a}$ could be tuned by changing drive wavelength
or precise tuning via ancillary laser and phase matching $\text{\ensuremath{\delta}}=\omega_{b}-2\omega_{a}$
can be tuned by environment~\citep{Hu2020}. To simulate the potential
interference effect, we fix the frequency of drive light to $100\,\mathrm{MHz}$
blue detuning from optimal drive SHG position and tune phase matching
$\text{\ensuremath{\delta}}$. As shown in Fig.$\,$\ref{FigS1},
The gain effect form $G_{a}$ cascaded term can be restrained by tuning
the value of $\delta$ but the total conversion efficiency still keeps
at high level at the same time.

\subsection{Parameter dependence of conversion efficiency}

In our theoretical model, the intensity of SFG output signal is mainly
determined by the frequency detunings, the coupling strengths $g_{2,3}$,
and also the cavity loss rates $\kappa_{a(b)1}$ and $\kappa_{a(b)0}$.
To provide an intuitive picture of the conversion, we numerically
investigated the parameter dependence of the conversion efficiencies
at different conditions. During the simulation, we consider the actual
situation in which we can only put the drive laser on the blue detuning
with respect to the resonance because of the thermal bistability~\citep{Hu2020}.
Besides, we change the phase matching $\text{\ensuremath{\delta}}=\omega_{b}-2\omega_{a}$
and the drive frequency to change $\Delta_{a}$ in the simulation
to find the optimal condition to achieve high conversion efficiency
or low noise. 
\begin{figure}[h]
\begin{centering}
\includegraphics[width=0.7\textwidth]{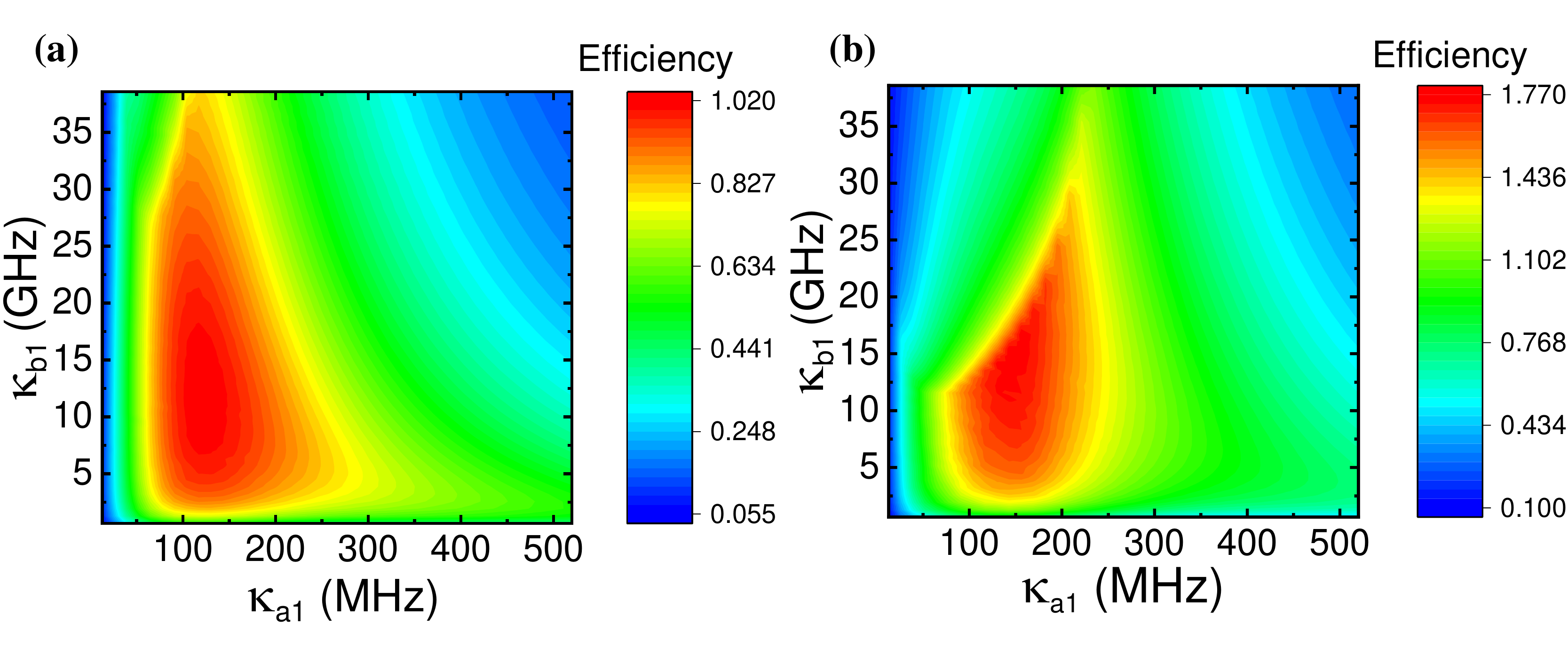}
\par\end{centering}
\caption{Simulation of parameter dependence between conversion efficiency and
external coupling rate $k_{a1},k_{b1}$. (a) The drive power is $50\,\mathrm{mW}$.
(b) The drive power is $80\,\mathrm{mW}$.}

\label{FigS2}
\end{figure}

For the influence of nonlinear coupling strength, the conversion could
be improved by increasing the coupling strengths, as shown in Fig.$\,$\ref{Fig4}(c)
in the main text. Figure$\,$\ref{FigS2} summarized the results for
various $\kappa_{a,1}$ and $\kappa_{b,1}$ parameters, for drive
strength of $50$ and $80\,\mathrm{mW}$, respectively. Here, we fixed
the intrinsic losses $\kappa_{a,0}$ and $\kappa_{b,0}$ to the experimental
parameters, because the intrinsic losses are mainly determined by
the material and fabrications, while the external losses could be
varied easily by designing the microring coupler geometry. The results
have shown that the optimal conversion could be achieved for a moderate
external coupling loss, because of the balancing of the external drive
and signal input to modes, nonlinear conversion interaction cooperativity,
and the energy extraction from the output mode. Especially, the output
efficiency could significantly exceed unity due to the cascaded conversion
by $g_{2}$ and amplification effect by $g_{3\mathrm{eff}}$, which
might play an important role in future applications.

\subsection{Noise in quantum frequency conversion}

In the previous discussions about the frequency conversion, we only
focus on the coherent frequency conversion efficiency and disregard
the potential noise generated by the amplification process. In our
current experiments, since the signal intensity is much larger than
the noise, we ignore the vacuum noise. However, for potential applications
of the degenerate SFG in quantum frequency conversion that signal
is on the single-photon level, we should also study the noise properties
of the device and also discuss how to suppress the noise in our scheme.

\begin{figure}[h]
\begin{centering}
\includegraphics[width=0.35\textwidth]{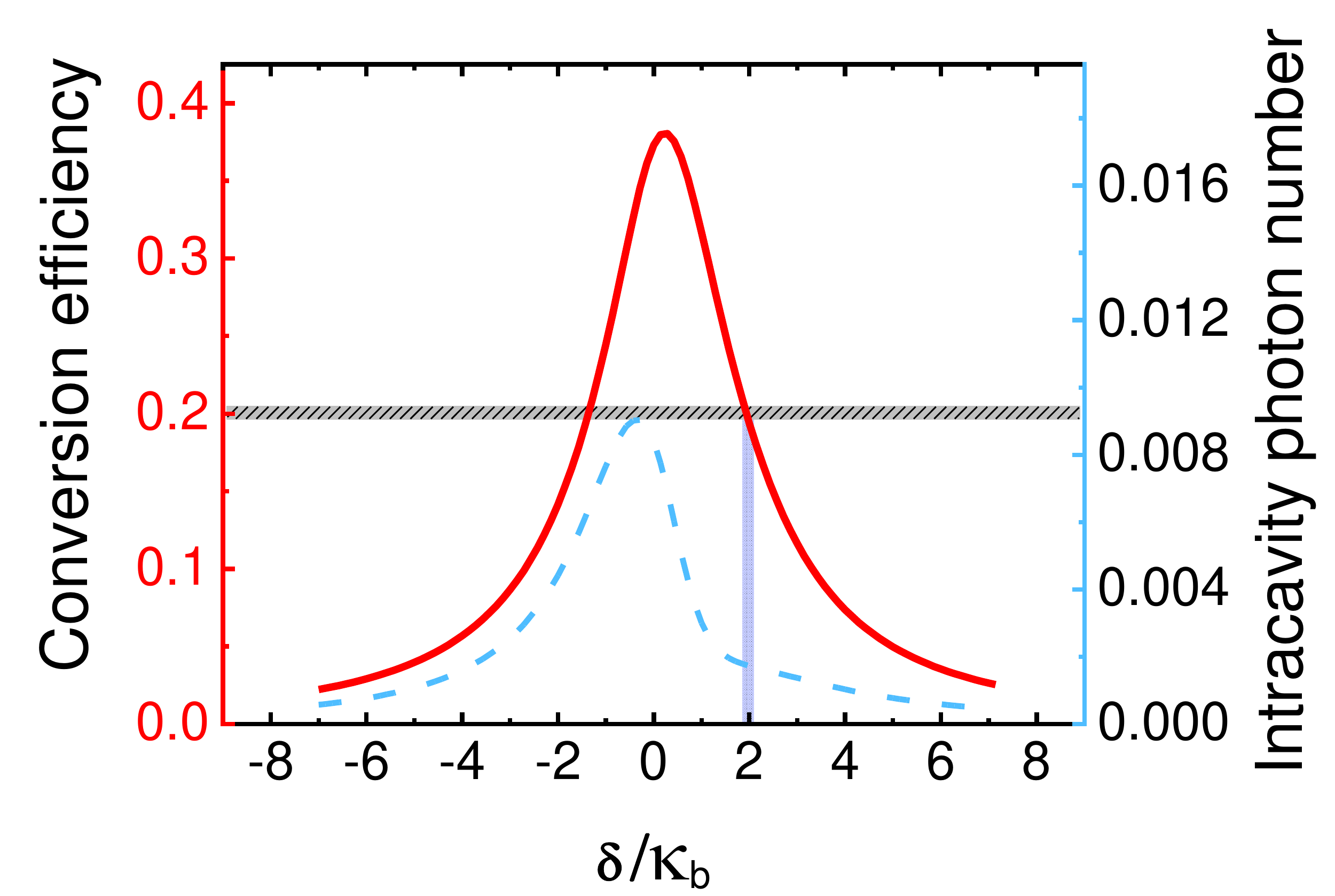}
\par\end{centering}
\caption{The relationship between $\delta$ and conversion efficiency and noise.
The parameters is same with our device in experiment. The on-chip
drive power is\textbf{ }$24\,\mathrm{mW}$. The red line represents
SFG conversion efficiency and the blue dash line represents the vacuum
noise.}

\label{FigS3}
\end{figure}

As shown in Eq.~(\ref{eq:sw}), the first two term are proportional
to signal light and indicate the conversion efficiency, and the remain
terms, i.e.\textbf{ $\frac{\delta\left(\omega\right)}{\text{\ensuremath{\pi}}}\left(\kappa_{a}F_{32}^{*}\left(\omega\right)F_{32}\left(\omega\right)+\kappa_{b}F_{34}^{*}\left(\omega\right)F_{34}\left(\omega\right)\right)$},\textbf{
}correspond to the amplified vacuum noise associating with the effective
Four-Wave Mixing (FWM) process. Shown in Fig.$\,$\ref{FigS3} is
the calculated conversion efficiency and noise photon number in the
cavity, with the parameters taken from our experimental device. When
the on-chip drive power is\textbf{ }$24\,\mathrm{mW}$, by tuning
the phase matching condition $\delta$ we can achieve $20\%$ conversion
efficiency with the average noise photon number in the cavity being
less than $0.002$. The results indicate that when the mean intracavity
signal photon is about $1$, our device can convert its frequency
with a probability higher than $20\%$, and the added noise is less
than $0.01$. It suggests that our device is capable to realize the
single-photon level quantum frequency conversion with considerably
high SNR. It is worth noting that the interference effect for the
two effective cascaded processes is beneficial for suppressing the
noise, by tuning the device to $\delta\sim2\kappa_{b}$ (the shadow
bar in Fig.~\ref{FigS3}).

\end{document}